\documentclass[sigconf,screen,review，10pt]{acmart} 
\usepackage{algorithmic}
\usepackage{graphicx}
\usepackage{textcomp}
\usepackage{multirow}
\usepackage{float}
\usepackage[utf8]{inputenc}
\usepackage[T1]{fontenc}
\usepackage{CJKutf8}
\usepackage{diagbox}
\usepackage{tabularx}
\usepackage{threeparttable}
\usepackage[normalem]{ulem}
\usepackage{mdframed}
\usepackage{tcolorbox}
\usepackage{hyperref}
\usepackage{url}
\usepackage{soul}
\usepackage{booktabs} 
\usepackage{array}    
\usepackage{comment}

\usepackage{enumitem}
\usepackage{balance}

\definecolor{lightgreen}{rgb}{0.67, 0.88, 0.69}  

\newboolean{showcomments}
\setboolean{showcomments}{true}

\ifthenelse{\boolean{showcomments}}
{\newcommand{\nbc}[3]{
 {\colorbox{#3}{\bfseries\sffamily\scriptsize\textcolor{white}{#1}}}
 {\textcolor{#3}{\sf\small$\blacktriangleright$\textit{#2}$\blacktriangleleft$}}
 }
}
{\newcommand{\nbc}[3]{}
 }

\newcommand\gopi[1]{\nbc{Gopi}{#1}{blue}}
\newcommand\vivi[2]{\nbc{vivi}{#1}{brown}}
\newcommand\ahmed[3]{\nbc{Ahmed}{#1}{cyan}}


\begin{document}
\renewcommand{\normalsize}{\fontsize{10}{12}\selectfont}
%
\title{LicenseGPT: A Fine-tuned Foundation Model for Publicly Available Dataset License Compliance}

\author{Jingwen Tan}
\affiliation{
  \institution{Sun Yat-Sen University}
  \country{China}
}
\email{tanjw9@mail2.sysu.edu.cn}

\author{Gopi Krishnan Rajbahadur, Zi Li, Xiangfu Song, Jianshan Lin}
\affiliation{
  \institution{ Huawei }
  \country{Canada, China}
}
\email{gopi.krishnan.rajbahadur1, lizi4, xiangfu.song1@huawei.com}

\author{Dan Li}
\authornote{Corresponding author}
\email{lidan263@mail.sysu.edu.cn}
\author{Zibin Zheng}
\email{zhzibin@mail.sysu.edu.cn}
\affiliation{
  \institution{Sun Yat-Sen University}
  \country{China}
}


\author{Ahmed E. Hassan}
\affiliation{
  \institution{Queen's University}
  \country{Canada}
}
\email{ahmed@cs.queensu.ca}
\renewcommand{\shortauthors}{Tan, et al.}

\begin{abstract}
Dataset license compliance is a critical yet complex aspect of developing commercial AI products, particularly with the increasing use of publicly available datasets. Ambiguities in dataset licenses pose significant legal risks, making it challenging even for software IP lawyers to accurately interpret rights and obligations. In this paper, we introduce LicenseGPT, a fine-tuned foundation model (FM) specifically designed for dataset license compliance analysis. We first evaluate existing legal FMs (i.e., FMs specialized in understanding and processing legal texts) and find that the best-performing model achieves a Prediction Agreement (PA) of only 43.75\%. LicenseGPT, fine-tuned on a curated dataset of 500 licenses annotated by legal experts, significantly improves PA to 64.30\%, outperforming both legal and general-purpose FMs. Through an A/B test and user study with software IP lawyers, we demonstrate that LicenseGPT reduces analysis time by 94.44\%, from 108 seconds to 6 seconds per license, without compromising accuracy. Software IP lawyers perceive LicenseGPT as a valuable supplementary tool that enhances efficiency while acknowledging the need for human oversight in complex cases. Our work underscores the potential of specialized AI tools in legal practice and offers a publicly available resource for practitioners and researchers. Moreover, LicenseGPT has the potential to assist AI software developers in managing preliminary license checks before involving legal counsel, helping to avoid costly late-stage rework and ensuring AI software compliance.
\end{abstract}



\acmYear{2025}\copyrightyear{2025}
\setcopyright{acmlicensed}
\acmConference[FSE Companion '25]{33rd ACM International Conference on the Foundations of Software Engineering}{June 23--28, 2025}{Trondheim, Norway}
\acmBooktitle{33rd ACM International Conference on the Foundations of Software Engineering (FSE Companion '25), June 23--28, 2025, Trondheim, Norway}
\acmDOI{10.1145/3696630.3728530}
\acmISBN{979-8-4007-1276-0/25/06}

\maketitle

\vspace{-0.5cm}


\section{Introduction}
\label{Sec:Introduction}

AI-powered software, particularly Foundation Models (FMs) like GPT and LLaMa, is growing rapidly and powering commercial applications such as GitHub Copilot~\cite{githubcopilot2021} and ChatGPT-4~\cite{openai2023gpt4}. However, building AI-powered software involves more than just sophisticated AI models—datasets are crucial throughout the development lifecycle. Consider the commercial software engineering lifecycle proposed for AI-powered software by Amershi~\textit{et~al}.~\cite{amershi2019software}. They cover stages such as data collection, cleaning, and model training, yet they do not include legal compliance as a part of it, a critical oversight that can lead to significant legal risks.

Despite the critical role of datasets, considerations regarding their licenses in AI-powered software development are often neglected. AI models are typically trained on large volumes of publicly available datasets~\cite{hassan2024rethinking, Gopi2021can, benjamin2019towards}. For example, GPT-3.5 processed 45~TB of public data, filtering it to 570~GB for training~\cite{brown2020language}. 

However, many publicly available datasets lack clear licenses, leaving rights and obligations uncertain, particularly concerning commercial use. Longpree~\textit{et al.}~\cite{longpre2023data} shows that over 66\% of dataset licenses are misrepresented, often with more permissive terms than intended. Several studies also found widespread misrepresentations stemming from unclear or convoluted license terms, posing legal risks~\cite{benjamin2019towards,Gopi2021can}. Recent lawsuits against companies like Google and OpenAI~\cite{HarvardGazette2023OpenAI, VOANews2023NYTimesSues} underscore the critical need for accurate dataset license compliance, particularly in commercial settings.

However, stringent compliance with dataset licenses is challenging. These licenses define dataset user's rights and obligations, determining whether datasets can be used for commercial applications or redistribution, and act as software requirements throughout the AI-powered software development lifecycle. Unfortunately, dataset licenses often lack standardized formats and detailed stipulations, posing significant impediments even for software IP lawyers. This burden is often shouldered by developers as well, particularly in smaller organizations and fast-paced AI software development environments where specialized legal counsel may be scarce. Early awareness of license constraints can avert costly late-stage rework. Publicly available dataset licenses frequently lack the clarity found in open-source software (OSS) licenses. For example, the CIFAR-10 license merely requests citation without specifying rights like the permissibility of using the data for training commercial AI models, while the ImageNet license restricts commercial use but remains ambiguous on the specific conditions under which the data can be used for non-commercial purposes. As Benjamin~\textit{et~al.}~\cite{benjamin2019towards} argue, this ambiguity allows for creative interpretations, such as commercializing the outputs of models trained on ImageNet since the dataset itself isn't used directly in a commercial context. 

Moreover, datasets like ImageNet and CIFAR-10 are compiled from various sources each with its own different licenses, complicating the determination of the overall dataset license. Dataset creators often fail to document original source licenses or consider their impact on the aggregated dataset's license, leading to unclear or potentially unlawful licenses and exposing consumers to risks.

In this paper, we shed light on and address the critical challenge of identifying the rights and obligations within a dataset's license. Identifying the rights and obligations with a dataset's license is a crucial yet laborious task. The complexity arises from the unique challenges posed by the different categories of dataset licenses commonly used in publicly available datasets: General Licenses, Customized Licenses, and Official Terms of Use or Service. General Licenses, often adapted from OSS formats, present non-straightforward obligations when applied to datasets. Customized Licenses contain highly specific, context-dependent clauses that add complexity. Official Terms of Use or Service are characterized by dense, legally nuanced language and complex technical jargon. Although Montreal Data Licenses~\cite{benjamin2019towards} and RAIL licenses~\cite{szpyt2020RAIL} aim to clarify these issues, they have not been widely adopted, and providers continue to use custom licenses with ambiguous terms. Rajbahadur~\textit{et~al.}~\cite{Gopi2021can} highlight the need for a systematic and transparent license interpretation tool with expert-in-the-loop capabilities to empower software engineers and IP lawyers in efficiently interpreting and assessing dataset licenses. 

Recently, FMs, particularly, Large Language Models (LLMs) have demonstrated impressive capabilities in text processing~\cite{zhao2023survey}, including legal documents. Several legal FMs have been developed to comprehend legal literature, primarily offering question-and-answer services~\cite{LawGPTzh,huang2023lawyerLLaMA,cui2023chatlaw,fuzi.mingcha,LexiLaw,HanFei,zhihai,songLawGPT}. However, these FMs focus on general legal texts, whereas dataset license compliance demands a nuanced understanding of license-specific terminology and context. Effective compliance requires the FM to grasp context-specific conditions, especially with custom licenses containing unique terms.

To address these specialized needs, we propose \textbf{LicenseGPT}, a fine-tuned FM specifically tailored for dataset license compliance analysis. We fine-tuned LicenseGPT using a Dataset Licenses (DL) dataset, comprising 500 publicly available dataset licenses collected from platforms like Hugging Face and GitHub, annotated by software IP lawyers. Each license is labeled to indicate whether it permits commercial use, prohibits it, or has ambiguous terms, along with the underlying reasons, associated rights, and obligations. Through the following research questions, we aim to assess the performance and effectiveness of LicenseGPT in improving dataset license compliance analysis compared to existing legal and general-purpose FMs:

\begin{itemize}[leftmargin=*]
    \item \textbf{RQ1: How effectively do current legal FMs perform on the task of dataset license compliance analysis?}
    
    \textit{Results:} The best-performing legal FM, LawGPT, achieves a Prediction Agreement (PA) score of 43.75\%, outperforming other legal FMs but with a moderate Semantic Similarity (SS) score of 50.25\%. General-purpose FMs like ChatGPT-4 achieve high SS scores but low PA scores, indicating that they produce semantically similar but inaccurate responses in this context.
    
    \item \textbf{RQ2: Does LicenseGPT enhance the accuracy of dataset license compliance analysis compared to existing legal and general-purpose FMs?}
    
    \textit{Results:} LicenseGPT achieves a PA score of 64.30\%, surpassing LawGPT by 20.55\% and the best-performing general-purpose FM, Qwen-1.5, by 4.58\%. This improvement is statistically significant with a large effect size. LicenseGPT also attains a higher SS score of 85.80\%, surpassing LawGPT by 35.55\%, and Qwen-1.5 by 1\%, indicating improved alignment with expert responses.
    
    \item \textbf{RQ3: How do software IP lawyers perceive the usefulness of LicenseGPT in dataset license compliance analysis?}
    
    \textit{Results:} Software IP lawyers found LicenseGPT valuable in practice. Lawyers using LicenseGPT completed analyses in an average of 6 seconds per license, compared to 108 seconds without it, which is a 94.44\% reduction in time. While they appreciated the efficiency gains, they also noted the need for careful validation in complex cases and recognized LicenseGPT as a valuable supplementary tool to be integrated into their workflows.
\end{itemize}

\noindent To support the community and encourage further research, we have open-sourced LicenseGPT~\cite{LicenseGPT} model (not the dataset). We also present several recommendations with actionable steps for both software engineering practitioners and researchers. Below, we list the contributions of our paper.

\begin{itemize}[leftmargin=*]
    \item We evaluate existing legal FMs on the task of dataset license compliance analysis and identify their limitations.
    \item We develop LicenseGPT, a fine-tuned FM with a significantly improved accuracy in interpreting dataset licenses.
    \item We conduct a user study with software IP lawyers to assess the practical utility of LicenseGPT in legal workflows.
    \item We open-source LicenseGPT to support the community and encourage further research in this critical area~\cite{LicenseGPT}.
    \item We identify immediate challenges in integrating dataset license compliance into AI software engineering lifecycle and provide recommendations to address these issues.
    
\end{itemize}

In addition to assisting software IP lawyers in expediting dataset license compliance analysis, LicenseGPT also serves as a valuable tool for developers managing preliminary license checks before involving legal counsel. By providing timely and accurate guidance on dataset constraints, LicenseGPT fosters effective collaboration between technical and legal teams, prevents costly late-stage rework, and enhances the efficiency of the AI software development lifecycle.

\subsection{Paper organization}
Section \ref{sec:background} covers the Background and Related Work, including legal protections, compliance challenges, and legal foundation models. Section \ref{sec:studydesign} describes the Study Design, including our dataset DL, experiment setup, and research questions (RQ1, RQ2, RQ3). Section \ref{sec:results} presents the results for each research question. Section \ref{sec:disscuss} discusses our findings in detail. In Section \ref{Sec:Threat}, we address the threats to validity. Finally, Section \ref{sec:conclusion} highlights the importance of integrating legal compliance into the AI software engineering lifecycle to enhance software quality and reduce legal risks, emphasizing the need for tools like LicenseGPT alongside human oversight.

\section{Background and Related Work}
\label{sec:background}
\subsection{Legal Protections Related to Datasets}

Datasets used in commercial contexts are governed primarily by copyright and contract law, providing significant safeguards across jurisdictions despite specific variations.

\smallskip\noindent\textit{Copyright Law} protects creative works from unauthorized use, including copying or reproduction without explicit permission from the copyright holder~\cite{canadacopyright}. Data contained in publicly available datasets may be copyright-protected, and unauthorized commercial use can constitute infringement~\cite{benjamin2019towards,peng2021mitigating}. While exceptions like the Fair Use doctrine in the United States permit certain uses without permission if they do not cause material harm to the copyright holder, as seen in \textit{Authors Guild v. Google}~\cite{lawsuit,USFairUse}, other jurisdictions like the UK, Canada, and the EU have stricter regulations. In these regions, fair dealing exceptions and directives like the EU's Text and Data Mining Directive typically restrict such uses to non-commercial purposes without explicit consent~\cite{CanadaFairDealing,triaille2014study}. Consequently, using datasets with copyrighted content for commercial AI-powered software development can lead to legal challenges depending on the jurisdiction.

\smallskip\noindent\textit{Contract Law} governs the agreements under which copyrighted materials are licensed. Copyright holders can issue licenses detailing the granted rights and required obligations for use. Violating these terms may result in a breach of contract. The precedence of copyright law versus contract law varies by jurisdiction, but contract law enables dataset licenses to permit commercial use without infringing copyright.

\subsection{Challenges in Dataset License Compliance} 

The primary goal in evaluating dataset licenses for commercial use is to determine whether a dataset can be utilized in specific scenarios, such as model training or redistribution, while ensuring compliance with the license terms. Compliance is crucial globally, serving as both a functional and non-functional requirement for AI-powered software~\cite{ingolfo2013arguing,breaux2008analyzing,kiyavitskaya2008automating,zeni2015gaiust}. When AI-powered software uses publicly available datasets, it implicitly enters into an agreement with the copyright holders, necessitating adherence to the rights and obligations outlined in the license. Failure to comply can result in serious legal risks~\cite{ingolfo2013arguing,breaux2008analyzing,kiyavitskaya2008automating,zeni2015gaiust}.

However, publicly available dataset licenses often lack clarity regarding usage rights and obligations, making compliance challenging for software engineers~\cite{benjamin2019towards}. This ambiguity complicates the process of translating license obligations into software requirements, which is critical for maintaining compliance. In situations where legal requirements are unclear, due diligence is essential to avoid breaches~\cite{breaux2008analyzing,kiyavitskaya2008automating}. Software engineers must trace rights and obligations from licenses to software requirements, documenting and justifying their interpretations and implementations~\cite{breaux2008analyzing}. Therefore, it is vital for software engineers and software IP lawyers to accurately identify the rights and obligations associated with publicly available datasets.

Although initiatives like the Montreal Data License~\cite{benjamin2019towards}, RAIL licenses~\cite{szpyt2020RAIL}, and dataset-specific licenses like PDDL~\cite{pddl2018} and CC BY~\cite{ccby2013} aim to clarify these issues, they have not been widely adopted. Dataset providers frequently use custom licenses with ambiguous terms, complicating the identification of rights and obligations. Recent studies show that over 66\% of publicly available dataset licenses are misrepresented on public platforms, often with more permissive terms than intended by the authors~\cite{longpre2023data}. This misrepresentation is prevalent due to unclear or convoluted license specifications~\cite{benjamin2019towards,Gopi2021can}. Additionally, the complexity of data ecosystems, where datasets are built upon other datasets with various sources and licenses, makes tracking data usage and understanding contributions increasingly difficult~\cite{barclay2019towards}. This complexity poses significant challenges in ensuring compliance in commercial settings. These challenges highlight the need for automated tools that can assist in accurately interpreting dataset licenses, particularly in the context of commercial AI-powered software development.

\subsection{Open Source License Compliance}

The enforcement of open-source licenses under copyright law was established in the landmark case \textit{Jacobsen v. Katzer} (Fed. Cir. 2008)~\cite{jacobsen2008}, where violating the terms of an open-source license was ruled to constitute copyright infringement. This set a legal precedent legitimizing and protecting the open-source movement~\cite{btlj2015}.

Following the ruling in \textit{Jacobsen v. Katzer}, the rise of Open Source Software (OSS) increased attention on license compliance, especially as OSS reuse became prevalent in software development. Researchers and practitioners developed tools and methodologies to detect and resolve license violations~\cite{16-german2012method,17-german2009license,27-kapitsaki2017automating,51-van2014tracing,56-zhang2010automatic,wsgr2017,btlj2015}. Commercial and open-source tools like BlackDuck~\cite{blackduck_website} and FOSSology~\cite{jaeger2009fossology,hansen2010fossology} are widely used to identify OSS licenses and ensure compliance with intended licensing frameworks. Resources like the Open Source Initiative (OSI)~\cite{osi}, GitHub's licensing guide~\cite{github_guide}, and TLDRLegal~\cite{tldrlegal} help practitioners understand the rights and obligations of various OSS licenses.

However, methods for OSS license compliance cannot be directly applied to dataset licenses. Publicly available dataset licenses often contain unclear and ambiguous terms regarding commercial use~\cite{benjamin2019towards,Gopi2021can}. Therefore, automated approaches for identifying rights and obligations for dataset licenses are needed, and our study addresses this gap.

\subsection{Ensuring legal compliance in commercial software}

Ensuring compliance with legal and licensing requirements is imperative for AI-powered software, as it directly influences user trust and legal viability. While organizations have established Open Source Program Offices (OSPOs)~\cite{munir2021OSPO} to oversee open-source compliance and governance for traditional software, these offices often aren't equipped to handle the distinct challenges presented by AI-powered software—particularly those related to dataset and model licensing. Consequently, initiatives like OpenChain~\cite{OpenChainProject} (ISO 5230 and 18974 standards on open source license and security compliance) have only recently initiated efforts to address these issues by establishing an AI study group~\cite{OpenChainAIWorkshop2024}. Our paper bridges this gap by highlighting the complexities of dataset license compliance that OSPOs~\cite{munir2021OSPO} must navigate when releasing AI-powered software. By introducing LicenseGPT, we offer an automated solution to streamline compliance processes, enabling organizations to meet licensing obligations while upholding software quality and reducing legal risks. 

\subsection{Legal Foundation Models}
\label{sec:subsec:legal FMs}
Developing legal FMs involves several key steps. First, we select a suitable base model, such as LLaMA~\cite{touvron2023llama}, Chinese-LLaMA~\cite{cui2023chinesellama}, or ChatGLM~\cite{du2022glm,zeng2022glm}. Choosing a robust FM reduces the additional training required for high performance in legal tasks and ensures higher accuracy in a legal context. Next, we perform continual pretraining to adapt the FM to the legal domain by training on domain-specific data like legal statutes and case law, enriching the model's understanding of legal language and concepts. Then, we apply instruction fine-tuning, refining the FM with specific instructions tailored to legal tasks to ensure it generates outputs that are accurate and aligned with legal professionals' expectations~\cite{lin2024legal}. Finally, we focus on prompt design, carefully crafting inputs including, system prompts, user prompts along with examples to guide the model's output, enabling it to produce relevant and accurate responses in the complex legal domain.

Following these steps, various legal FMs have emerged with unique capabilities tailored to distinct legal contexts. For example, LawGPT\_zh (6B)~\cite{LawGPTzh} and Lawyer LLaMA (13B)~\cite{huang2023lawyerLLaMA} enhance legal text processing, while ChatLaw (13B)~\cite{cui2023chatlaw} integrates external knowledge bases to reduce misinformation. Fuzi.mingcha (6B)~\cite{fuzi.mingcha} and LexiLaw (6B)~\cite{LexiLaw} are developed for case analysis and legal consulting within the Chinese legal framework. HanFei (7B)~\cite{HanFei} and Wisdom Interrogatory (7B)~\cite{zhihai} focus on integrating intelligent legal systems into judicial practices. Despite these advancements, existing legal FMs are not tailored to dataset license compliance, a specialized challenge requiring additional fine-tuning and customization. Our study fills this gap by developing \textbf{LicenseGPT}, a model that is specifically designed for dataset license compliance analysis.

\section{Study Design}
\label{sec:studydesign}

\begin{figure*}[h]
  \centering
  \includegraphics[width=2.0\columnwidth]{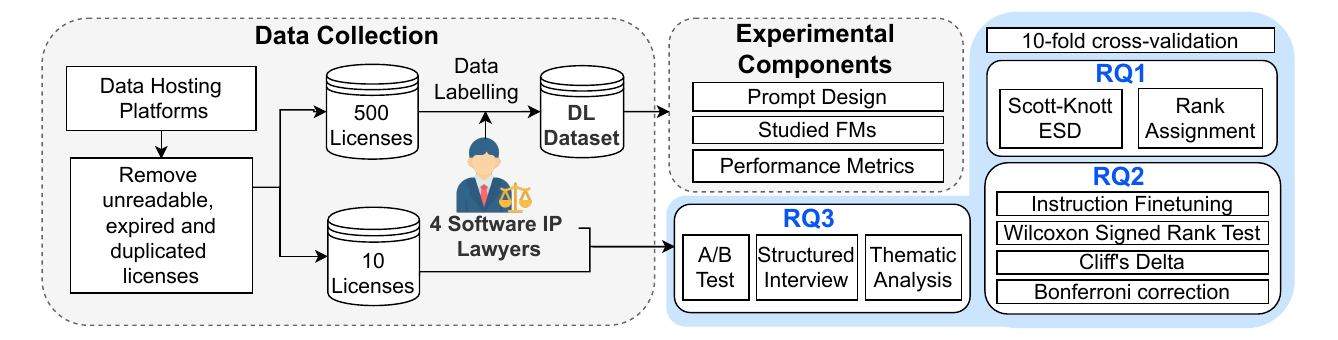}
  \caption{Overview of our study design}
  \label{fig:overview}
\end{figure*}


\subsection{Dataset Licenses (DL) Dataset}
In this section, we describe the creation of the dataset that we use in our study to understand the capabilities of the exisiting legal FMs on the task of data license compliance and creating LicenseGPT. Figure~\ref{fig:overview} presents an overview of our study design including our DL dataset creation process. To create LicenseGPT we required a dataset that includes commonly used publicly available licenses, annotated with information on whether the license permits commercial use, as well as the specific rights and obligations it outlines. Since no such dataset currently exists, we built one from scratch. We detail the steps that we undertook to create our DL dataset below.

\smallskip\noindent\textbf{Step 1: Dataset License Collection.} We first collected 37, 204 dataset licenses that are associated with publicly available datasets hosted in 10 common machine learning dataset hosting platforms. We collected all the licenses associated with the publicly available datasets hosted on Hugging Face~\cite{huggingface}, Google Cloud~\cite{googlecloud}, Kaggle~\cite{kaggle}, Zenodo~\cite{zenodo}, GitHub~\cite{github}, GitLab~\cite{gitlab}, Microsoft Azure~\cite{microsoftazure}, DataHub~\cite{datahub}, Amazon S3~\cite{amazons3}, and Figshare~\cite{figshare}. We also hosted all the collected licenses on the OpenDataology open source project website~\cite{opendataology}. It is important to note that the collected licenses encompass all three categories described in Section~\ref{Sec:Introduction}: General licenses, Customized licenses, and Official terms of use. Table~\ref{tab:LawyerAdvise} presents the number of licenses from each category contained in our dataset.

\smallskip\noindent\textbf{Step 2: Dataset filtering.} We removed the unreadable, expired and duplicated licenses. After such filtering, we were left with 500 valid dataset licenses.

\begin{table}[!htbp]
\fontsize{10}{12}
\caption{Examples of Lawyer's Advice for Different Types of Licenses}
\centering
\scriptsize
\label{tab:LawyerAdvise}

\begin{tabular}{@{}lr@{}}
\toprule
\textbf{License Type} & \textbf{Number of licenses} \\ \midrule
\textbf{General License} & 146 \\
\textbf{Customized License} & 186 \\
\textbf{Official Terms of Use or Service} & 168 \\
\midrule
\textbf{Total} & 500 \\
\bottomrule
\end{tabular}

\end{table}




\smallskip\noindent\textbf{Step 3: Dataset labelling.} We enlisted four software IP lawyers from our company, each with 8, 10, 15, and 20 years of experience, to label each dataset license. The software IP lawyers were instructed to categorize the licenses into three labels: ``allows commercial use,'' ``does not allow commercial use,'' and ``unclear if the license allows commercial use''. Additionally, for licenses that permit commercial use, the lawyers were asked to document the specific rights and obligations associated with the license. For the other two categories, they were instructed to provide reasons for their classifications.

Each lawyer independently labeled the dataset licenses and provided their recommendations. Following this, we held a meeting where all the lawyers who labeled the licenses discussed their labels and advice. Through this collaborative discussion, we arrived at the final labels and associated guidance for each license. An example label and recommendation provided by the lawyers for CIFAR-10 dataset's license look like \textit{\textbf{label}: unclear if the license allows commercial use, \textbf{recommendation}:  The provided information suggests referencing a technical report when using the dataset, which details its collection methodology. However, it doesn't specify if commercial use is permitted; such permission usually requires explicit mention in the use agreement or additional licensing. To confirm commercial use rights, one should review the full agreement or consult the dataset's creator.} 

\subsection{Prompt Design}
For FMs, prompt design is essential for guiding models to generate relevant and accurate responses~\cite{hassan2024rethinking}. Effective prompts not only improve model performance but also ensure precision and adaptability across the given task. In this study, we developed a structured prompt design for both \textbf{LicenseGPT} and the evaluation of existing legal FMs for dataset license comprehension. Following the advice of Hassan~\textit{et al.}~\cite{hassan2024rethinking}, we seperated system and user prompt. We follow this modular design to generate legally sound, focused responses, enhancing the models' ability to perform nuanced legal analyses. We provide both the system prompt and user prompt that we used in our online repository~\cite{LicenseGPT}.



\textbf{System Prompt Design.} System prompts define the model’s role and scope. By setting the model to act as a software IP lawyer, we ensure its responses adhere to legal standards. This approach sharpens the model’s focus and steers it towards relevant legal expertise. \textit{Role Assumption:} We assign the model the role of a software IP lawyer, ensuring its responses are legally sound. Defining roles enhances task relevance in specialized domains \cite{LegalBench}. \textit{Task Definition:} We clearly define the task for the model, instructing it to assess whether a dataset can be used commercially, thus maintaining a focused objective \cite{brown2020language}. \textit{Focus Specification:} We direct the model to concentrate on the legality of commercial use, reducing distractions and increasing output accuracy \cite{patil2024review}. 

\textbf{User Prompt Design.} User prompts provide step-by-step guidance on how the model should approach the task. Clear user prompts enhance model performance by offering specific direction. \textit{Context:} We provide clear context by specifying whether the task pertains to a dataset license (both generalized and customized) or a website’s usage agreement. The actual license content is also given as context to ensure relevance \cite{LegalBench,getindata2023}. \textit{Query:} We instruct the model to determine if the dataset's license allows commercial use and what obligations arise, keeping the focus on the core legal issue \cite{yadav2023scaling}. \textit{Logic:} We guide the model to provide legally reasoned responses and acknowledge uncertainties, which improves transparency \cite{yadav2023scaling}.

\subsection{Zero-Shot Experimentation}
\label{sec:zeroshot}
We conducted all experiments under a zero-shot setting, where FMs generate responses without prior examples or task-specific fine-tuning during inference~\cite{neptuneai_zeroshot,huggingface_zeroshot}. We chose this approach to ensure that LicenseGPT remains accessible to small companies and academic institutions that may lack proprietary legal analyses even for examples. By focusing on zero-shot performance, we aim to create an open-source tool that can be readily adopted without the need for additional resources or data, making it practical for widespread use.

\subsection{Studied FMs}
In our study, we selected eight recent state-of-the-art legal FMs (discussed in Section~\ref{sec:subsec:legal FMs}) with parameter sizes ranging from 6 billion to 13 billion. Since all the legal FMs required local hosting, we limited our selection to models with up to 13 billion parameters, excluding larger models. Notably, while LawGPT\_zh and LawGPT share similar names, they are distinct models: LawGPT\_zh is fine-tuned from ChatGLM-6B using LoRA 16-bit instruction tuning, whereas LawGPT is based on Chinese-LLaMA and underwent legal vocabulary expansion, large-scale pre-training on a legal corpus, and subsequent instruction fine-tuning.

Additionally, we examined 3 state-of-the-art, general-purpose chat-tuned FMs, including ChatGPT-4, LLaMa-2, and Qwen-1.5, which we accessed through third-party APIs, to evaluate their performance in dataset license compliance analysis. Table~\ref{tab:StudiedFMs} lists the studied models along with their parameter sizes.

\begin{table}[!htbp]
\fontsize{10}{12}
\centering
\caption{Information of Studied FMs}
\scriptsize
\label{tab:StudiedFMs}
\begin{tabular}{>{\raggedright}p{4cm} >{\raggedright}p{2cm} >{\raggedleft\arraybackslash}p{1cm}}
\toprule
\centering\textbf{Model Name} & \centering\textbf{Base Model} & \textbf{Parameter} \\ \midrule
ChatGPT-4~\cite{openai2023gpt4} & GPT-4 & 175B \\ 
LLaMA-2~\cite{meta2023llama} & LLaMA-2 Base & 70B \\ 
Qwen-1.5~\cite{alibaba2023qwen} & Qwen Base & 110B \\ 
LawGPT\_zh~\cite{LawGPTzh} & ChatGLM & 6B \\ 
LawGPT~\cite{songLawGPT} & Chinese-Llama & 7B \\ 
Lawyer LLaMA~\cite{huang2023lawyerLLaMA} & LLaMA & 13B \\ 
ChatLaw~\cite{cui2023chatlaw} & LLaMA & 13B \\ 
fuzi.mingcha~\cite{fuzi.mingcha} & ChatGLM & 6B \\ 
LexiLaw~\cite{LexiLaw} & ChatGLM & 6B \\ 
HanFei~\cite{HanFei} & HanFei-1.0 & 7B \\ 
Wisdom Interrogatory~\cite{zhihai} & Baichuan & 7B \\ 
\bottomrule
\end{tabular}
\end{table}


\subsection{Studied Performance Metrics}

To evaluate the performance of the studied legal FMs and our proposed LicenseGPT we use the performance metrics in Table~\ref{tab:metrics}. Of the studied metrics, Prediction Agreement, Duplication Rate and Non-specific Response Rate are computed manually by three of the authors of this paper collaboratively by carefully analyzing each of the model's response and comparing it against the ground truth that is provided by the lawyers involved in the study.

\begin{table*}[!htbp]
\fontsize{10}{12}
\centering
\caption{Summary of Studied Performance Metrics}
\label{tab:metrics}
\scriptsize
\setlength{\tabcolsep}{10pt} 
\renewcommand{\arraystretch}{1.1} 
\begin{tabular}
{@{}p{1.2cm}p{4.5cm}l p{5cm}@{}}
\toprule
\textbf{Metric Name} & \textbf{Metric Description} & \textbf{Equation} & \textbf{Metric Interpretation} \\
\midrule

\textbf{Prediction Agreement (PA)} & Measures the alignment between model predictions and expert evaluations. Calculated as the percentage of correct predictions. & 

$\displaystyle \frac{\text{No. of Correct Predictions}}{\text{Total No. of Samples}} \times 100\%$
 & Higher PA means better model performance, crucial in legal contexts where precision is key. Low PA indicates divergence from expert evaluations. \\

\midrule

\textbf{Duplication Rate (DR)} & Calculates the percentage of repeated answers in the model’s output. & 
$\displaystyle \frac{\text{Number of Duplicate Answers}}{\text{Total Number of Answers}} \times 100\% $
 & High DR shows reliance on generic responses, problematic for nuanced legal tasks. Low DR indicates more context-specific answers. \\

\midrule

\textbf{Semantic Similarity (SS)} & Measures the semantic similarity between the model’s output and expert responses using BERT embeddings. & 
$\displaystyle \frac{\mathbf{X} \cdot \mathbf{Y}}{\|\mathbf{X}\| \|\mathbf{Y}\|}$
 & SS over 80\% suggests strong alignment with legal experts, ensuring contextually appropriate responses. \\

\midrule

\textbf{Non-Specific Response Rate (NRR)} & Calculates the percentage of instances where a model provides no specific judgment or uses a one-size-fits-all answer. & 
$\displaystyle \frac{\text{Number of Nonspecific Answers}}{\text{Total Number of Answers}} \times 100\% $
& High NRR shows overly general answers, while low NRR reflects the model's ability to offer detailed, context-aware judgments. \\

\midrule

\textbf{Average Response Speed (ARS)} & Calculates the average time taken for the model to generate an answer across all instances. & 
$\displaystyle \frac{\sum_{i=1}^{\text{\#Answers}} \text{Response time}_{i}}{\text{Total Number of Answers}}$
 & Lower ARS indicates faster, more efficient responses, which is beneficial for quick legal decision-making. \\

\bottomrule
\end{tabular}
\end{table*}

\section{Research Questions}
\label{sec:results}
In this Section, we present the approach and results of each of the studied RQs.
\begin{table*}[!htbp]
\scriptsize
\centering
\caption{Average performance and SK-ESD ranks of all the studied FMs}
\label{results:table1}
\begin{tabular}{@{}p{0.1cm} *{22}{p{0.1cm}}}
\toprule
& \multicolumn{2}{c}{\textbf{LG\_zh}} & \multicolumn{2}{c}{\textbf{F.ming}} & \multicolumn{2}{c}{\textbf{LexiLaw}} & \multicolumn{2}{c}{\textbf{HanFei}} & \multicolumn{2}{c}{\textbf{W.Int}} & \multicolumn{2}{c}{\textbf{LGPT}} & \multicolumn{2}{c}{\textbf{LLaMa}} & \multicolumn{2}{c}{\textbf{CLaw}} & \multicolumn{2}{c}{\textbf{CGPT4}} & \multicolumn{2}{c}{\textbf{LLM2}} & \multicolumn{2}{c}{\textbf{QWen-1.5}} \\
\midrule
& \multicolumn{1}{c}{\textbf{R}} & \multicolumn{1}{c}{\textbf{V}} & \multicolumn{1}{c}{\textbf{R}} & \multicolumn{1}{c}{\textbf{V}} & \multicolumn{1}{c}{\textbf{R}} & \multicolumn{1}{c}{\textbf{V}} & \multicolumn{1}{c}{\textbf{R}} & \multicolumn{1}{c}{\textbf{V}} & \multicolumn{1}{c}{\textbf{R}} & \multicolumn{1}{c}{\textbf{V}} & \multicolumn{1}{c}{\textbf{R}} & \multicolumn{1}{c}{\textbf{V}} & \multicolumn{1}{c}{\textbf{R}} & \multicolumn{1}{c}{\textbf{V}} & \multicolumn{1}{c}{\textbf{R}} & \multicolumn{1}{c}{\textbf{V}} & \multicolumn{1}{c}{\textbf{R}} & \multicolumn{1}{c}{\textbf{V}} & \multicolumn{1}{c}{\textbf{R}} & \multicolumn{1}{c}{\textbf{V}} & \multicolumn{1}{c}{\textbf{R}} & \multicolumn{1}{c}{\textbf{V}} \\
\midrule
\textbf{PA} & \multicolumn{1}{r}{6} & \multicolumn{1}{r}{35.71} & \multicolumn{1}{r}{7} & \multicolumn{1}{r}{30.95} & \multicolumn{1}{r}{4} & \multicolumn{1}{r}{40.71} & \multicolumn{1}{r}{8} & \multicolumn{1}{r}{22.05} & \multicolumn{1}{r}{3} & \multicolumn{1}{r}{43.02} & \multicolumn{1}{r}{2} & \multicolumn{1}{r}{43.75} & \multicolumn{1}{r}{9} & \multicolumn{1}{r}{19.05} & \multicolumn{1}{r}{9} & \multicolumn{1}{r}{19.05} & \multicolumn{1}{r}{11} & \multicolumn{1}{r}{18.06} & \multicolumn{1}{r}{5} & \multicolumn{1}{r}{40.28} & \multicolumn{1}{r}{\cellcolor{lightgreen}1} & \multicolumn{1}{r}{\cellcolor{lightgreen}59.72} \\
\textbf{DR} & \multicolumn{1}{r}{8} & \multicolumn{1}{r}{16.67} & \multicolumn{1}{r}{6} & \multicolumn{1}{r}{7.14} & \multicolumn{1}{r}{\cellcolor{lightgreen}1} & \multicolumn{1}{r}{\cellcolor{lightgreen}0} & \multicolumn{1}{r}{7} & \multicolumn{1}{r}{9.52} & \multicolumn{1}{r}{9} & \multicolumn{1}{r}{35.71} & \multicolumn{1}{r}{\cellcolor{lightgreen}1} & \multicolumn{1}{r}{\cellcolor{lightgreen}0} & \multicolumn{1}{r}{11} & \multicolumn{1}{r}{83.33} & \multicolumn{1}{r}{9} & \multicolumn{1}{r}{35.71} & \multicolumn{1}{r}{\cellcolor{lightgreen}1} & \multicolumn{1}{r}{\cellcolor{lightgreen}0} & \multicolumn{1}{r}{5} & \multicolumn{1}{r}{1.87} & \multicolumn{1}{r}{\cellcolor{lightgreen}1} & \multicolumn{1}{r}{\cellcolor{lightgreen}0} \\
\textbf{NRR} & \multicolumn{1}{r}{11} & \multicolumn{1}{r}{23.81} & \multicolumn{1}{r}{9} & \multicolumn{1}{r}{21.43} & \multicolumn{1}{r}{8} & \multicolumn{1}{r}{20.12} & \multicolumn{1}{r}{7} & \multicolumn{1}{r}{11.9} & \multicolumn{1}{r}{6} & \multicolumn{1}{r}{9.52} & \multicolumn{1}{r}{\cellcolor{lightgreen}1} & \multicolumn{1}{r}{\cellcolor{lightgreen}0} & \multicolumn{1}{r}{4} & \multicolumn{1}{r}{4.76} & \multicolumn{1}{r}{9} & \multicolumn{1}{r}{21.43} & \multicolumn{1}{r}{3} & \multicolumn{1}{r}{3.40} & \multicolumn{1}{r}{5} & \multicolumn{1}{r}{5.17} & \multicolumn{1}{r}{2} & \multicolumn{1}{r}{0.79} \\
\textbf{ARS} & \multicolumn{1}{r}{5} & \multicolumn{1}{r}{5} & \multicolumn{1}{r}{11} & \multicolumn{1}{r}{65} & \multicolumn{1}{r}{7} & \multicolumn{1}{r}{10} & \multicolumn{1}{r}{10} & \multicolumn{1}{r}{37} & \multicolumn{1}{r}{9} & \multicolumn{1}{r}{20} & \multicolumn{1}{r}{3} & \multicolumn{1}{r}{1.7} & \multicolumn{1}{r}{8} & \multicolumn{1}{r}{13} & \multicolumn{1}{r}{6} & \multicolumn{1}{r}{9} & \multicolumn{1}{r}{2} & \multicolumn{1}{r}{1.3} & \multicolumn{1}{r}{\cellcolor{lightgreen}1} & \multicolumn{1}{r}{\cellcolor{lightgreen}1.0} & \multicolumn{1}{r}{4} & \multicolumn{1}{r}{3.8} \\

\textbf{SS} & \multicolumn{1}{r}{11} & \multicolumn{1}{r}{31.78} & \multicolumn{1}{r}{7} & \multicolumn{1}{r}{44.39} & \multicolumn{1}{r}{8} & \multicolumn{1}{r}{39.06} & \multicolumn{1}{r}{9} & \multicolumn{1}{r}{37.01} & \multicolumn{1}{r}{10} & \multicolumn{1}{r}{34.36} & \multicolumn{1}{r}{6} & \multicolumn{1}{r}{50.25} & \multicolumn{1}{r}{4} & \multicolumn{1}{r}{65.45} & \multicolumn{1}{r}{4} & \multicolumn{1}{r}{63.38} & \multicolumn{1}{r}{1} & \multicolumn{1}{r}{94.80} & \multicolumn{1}{r}{2} & \multicolumn{1}{r}{92.00} & \multicolumn{1}{r}{3} & \multicolumn{1}{r}{83.10} \\

\textbf{Avg.} & \multicolumn{1}{r}{6} & \multicolumn{1}{r}{-} & \multicolumn{1}{r}{8} & \multicolumn{1}{r}{-} & \multicolumn{1}{r}{5.6} & \multicolumn{1}{r}{-} & \multicolumn{1}{r}{8.2} & \multicolumn{1}{r}{-} & \multicolumn{1}{r}{7.4} & \multicolumn{1}{r}{-} & \multicolumn{1}{r}{2.6} & \multicolumn{1}{r}{-} & \multicolumn{1}{r}{7.2} & \multicolumn{1}{r}{-} & \multicolumn{1}{r}{7.4} & \multicolumn{1}{r}{-} & \multicolumn{1}{r}{3.6} & \multicolumn{1}{r}{-} & \multicolumn{1}{r}{3.6} & \multicolumn{1}{r}{-} & \multicolumn{1}{r}{2.2} & \multicolumn{1}{r}{-} \\
\bottomrule

\end{tabular}
\begin{tablenotes}
   \scriptsize
   \item R- Rank; V- Value; LG\_zh (LawGPT\_zh), F.ming (fuzi.mingcha), W.Int (Wisdom Interrogatory), LGPT (LawGPT), LLaMa (Lawyer LLaMA), CGPT4 (ChatGPT-4), LLM2 (LLaMa-2).
 \end{tablenotes}
\end{table*}

\smallskip\noindent\textbf{\textit{RQ1: How effectively do current legal FMs perform on the task of dataset license compliance analysis?}}
\smallskip\noindent\textbf{Approach.} We collected all the studied performance metrics from each cross-validation run for the studied legal and general-purpose FMs. Specifically, we conducted 10-fold stratified cross-validation on a balanced 10\% subset of the DL dataset, ensuring representative coverage of commercial usability labels. We then applied Scott-Knott Effect Size Difference (SK-ESD)~\cite{ScottKnottESD}, similar to clustering, to rank the models based on these metrics. SK-ESD utilizes effect size, computed using Cohen’s $\Delta$~\cite{cohen1988statistical}, to group statistically similar models into the same rank. We chose SK-ESD since it is a non-parametric ranking method that yields statistically robust results~\cite{hassan2024rethinking}.

\smallskip\noindent\textbf{Results.~Result 1. LawGPT achieves a 43.75\% average Prediction Agreement (PA), outperforming all other studied legal FMs, but has a moderate Semantic Similarity (SS) score.} From Table~\ref{results:table1}, we observe that while LawGPT's PA score is slightly higher than that of Wisdom Interrogatory (43.75\% vs. 43.02\%) and LexiLaw (40.71\%), LawGPT has a lower SS score of 50.25\% compared to LexiLaw's 39.06\% and Wisdom Interrogatory's 34.36\%. This suggests that although LawGPT provides more accurate predictions, its responses are only moderately semantically similar to the expected answers provided by legal professionals.

Moreover, LawGPT has lower Duplication Rate (DR), Non-Specific Response Rate (NRR), and Average Response Speed (ARS) compared to other legal FMs. Specifically, LawGPT has a DR of 0\%, NRR of 0\%, and ARS of 1.7 seconds, whereas Wisdom Interrogatory has DR, NRR, and ARS scores of 35.71\%, 9.52\%, and 20 seconds, respectively. Lower DR and NRR indicate that LawGPT produces fewer duplicated and non-specific responses, enhancing its practical utility. However, the moderate SS score implies that while LawGPT's answers are accurate, they may differ in wording or style from the ground truth, highlighting an area for potential improvement in aligning its outputs more closely with expert responses.

\smallskip\noindent\textbf{Result 2. LawGPT outperforms two out of the three studied general-purpose FMs, including ChatGPT-4, in terms of Prediction Agreement (PA), but has a lower Semantic Similarity (SS) score.} Despite its strong performance on legal benchmarks~\cite{OpenAI2023}, ChatGPT-4 ranks last among all studied FMs in PA for dataset license compliance analysis, with a PA of 18.06\%, yet it achieves the highest SS score of 94.80\%. Similarly, LLaMA-2 has a PA of 40.28\% and an SS score of 92.00\%. Only Qwen-1.5 surpasses LawGPT in both PA (59.72\%) and SS (83.10\%) scores. In contrast, LawGPT has a higher PA of 43.75\% but a lower SS score of 50.25\%. This discrepancy indicates that general-purpose FMs like ChatGPT-4 and LLaMA-2 produce responses that are semantically similar to the expected answers but may lack accuracy in the specific context of dataset license compliance.

These results indicate the need for domain-specific training to achieve responses that are not only semantically similar but also accurate. LawGPT's higher PA score of 43.75\% demonstrates better alignment with expert judgments, even though its SS score is lower at 50.25\%, indicating that it prioritizes accuracy over similarity in wording or expression.


\smallskip\noindent\textbf{Result 3. LawGPT is the most suitable candidate FM for LicenseGPT.} LawGPT has the highest Scott-Knott Effect Size Difference (SK-ESD) rank across all metrics among the studied legal FMs, indicating superior overall performance. Furthermore, LawGPT can be locally fine-tuned without extensive computational resources, making it practical for organizations with limited capabilities. Since the Dataset License (DL) dataset is proprietary and subject to privacy concerns, using a model that allows local fine-tuning mitigates risks associated with third-party FM hosting. Therefore, LawGPT addresses both performance and practical considerations, making it the best candidate for developing LicenseGPT.

\smallskip\noindent\textbf{\textit{RQ2: Does LicenseGPT enhance the accuracy of dataset license compliance analysis compared to existing legal FMs?}}

\smallskip\noindent\textbf{Approach.} We selected the best-performing legal FM from RQ1 and fine-tuned it using the DL dataset to create LicenseGPT. Similar to RQ1, we conducted 10-fold stratified cross-validation, where during each run, we fine-tuned the selected FM on 9 folds using LoRA~\cite{zhang2023adaptiveLoRA}, known for its efficiency with minimal parameter tuning, ensuring computational efficiency. On average, each fine-tuning operation took 1200 seconds (i.e., 20 minutes). We then evaluated LicenseGPT's performance on the remaining fold.

To measure performance differences between LicenseGPT, the best-performing Legal FM from RQ1, and general-purpose FMs, we conducted a Wilcoxon signed-rank test, as it does not assume normality and is suited for paired data. To quantify the magnitude of performance differences, we applied Cliff's delta. The thresholds for interpreting Cliff's delta are: $0 < \Delta \leq 0.33$ indicates a small difference, $0.33 < \Delta \leq 0.66$ indicates a medium difference, and $\Delta > 0.66$ indicates a large difference. Additionally, we performed Bonferroni correction~\cite{dunn1961Bonferroni,rupert2012Bonferroni} due to the multiple pairwise comparisons across performance metrics.

\smallskip\noindent\textbf{Results.~Result 4. LicenseGPT significantly outperforms all studied general-purpose and legal FMs with a large effect size across all studied performance metrics.} LicenseGPT achieves a PA score of 64.30\%, surpassing LawGPT by 20.55\% and the best-performing general-purpose FM, Qwen-1.5, from RQ1, by 4.58\%. This improvement is statistically significant with a large effect size.

Furthermore, LicenseGPT attains an SS score of 85.80\%, higher than Qwen-1.5 (83.10\%) and LawGPT (50.25\%). While ChatGPT-4 and LLAMA-2 exhibit higher SS scores (94.80\% and 92.00\%, respectively), their low PA scores (18.06\% and 40.28\%) indicate that they often produce semantically similar but incorrect responses, reducing their reliability for accurate license compliance analysis.
\begin{table}[h]
\caption{Comparison of LicenseGPT's performance across the studied performance measures}
\label{tab:RQ2modelperformance}
\centering
\scriptsize
\begin{tabular}{cccccc}
\hline
\textbf{Model} & \textbf{PA(\%)} & \textbf{SS(\%)} & \textbf{DR(\%)} & \textbf{NRR(\%)} & \textbf{ARS (s)} \\ \hline
ChatGPT4-175B & \cellcolor{yellow!60}\textbf{18.06} & \cellcolor{yellow!60}\textbf{94.80} & \cellcolor{yellow!60}\textbf{0} & 3.40 & \textbf{1.30} \\
LLAMA2-70B & \cellcolor{yellow!60}\textbf{40.28} & \cellcolor{yellow!60}\textbf{92.00} & 1.87 & \cellcolor{yellow!60}\textbf{5.17} & \textbf{1.00} \\
Qwen1.5-110B & \cellcolor{yellow!60}\textbf{59.72} & \cellcolor{yellow!60}\textbf{83.10} & \cellcolor{yellow!60}\textbf{0} & 0.79 & \cellcolor{yellow!60}3.80 \\ 
LawGPT-7B & \cellcolor{yellow!60}\textbf{43.75} & \cellcolor{yellow!60}\textbf{50.25} & \cellcolor{yellow!60}\textbf{0} & \cellcolor{yellow!60}\textbf{0} & \textbf{1.7} \\ \hline
LicenseGPT & 64.30 & 85.80 & 5.71 & 3.4 & 2.40 \\ \hline
\end{tabular}
\begin{tablenotes}
    \item Statistically significant results are indicated with bold font.
    \item LicenseGPT has a large effect size with yellow background color.
\end{tablenotes}
\end{table}

In terms of DR and NRR, LicenseGPT maintains a competitive performance with DR of 5.71\% and NRR of 3.4\%, which are acceptable for practical applications. Although LawGPT has a DR and NRR of 0\%, its lower PA and SS scores suggest less accurate and less context-specific responses. LicenseGPT's Average Response Speed (ARS) is 2.40 seconds, slightly higher than LawGPT (1.7 seconds) but still within a practical range for user interaction.

To illustrate the qualitative improvements, Table~\ref{tab:RQ2Eample} presents example responses from LicenseGPT and other models when interpreting the commercial usability of a dataset licensed under CC BY-NC 4.0. LicenseGPT provides a detailed and accurate analysis, clearly explaining the restrictions and offering actionable guidance, whereas other models provide vague or incomplete responses.

These results demonstrate that LicenseGPT enhances the accuracy and reliability of dataset license compliance analysis compared to existing legal and general-purpose FMs, making it a valuable tool for practitioners. However, despite these improvements, a PA of 64.30\% indicates that there is still room for further enhancement in model accuracy.

\begin{table*}[ht]
\fontsize{10}{12}
\centering
\scriptsize
\caption{Comparison of Model Outputs for Dataset Licensed under CC BY-NC 4.0}

\label{tab:RQ2Eample}
\resizebox{\textwidth}{!}{%
\begin{tabular}{p{15cm}}
\toprule
\textbf{Input: Can I use a dataset licensed under CC BY-NC 4.0 in my commercial project? What are the restrictions, if any?} \\
\textcolor{gray}{\footnotesize{License Text: CC BY-NC 4.0 stands for Creative Commons Attribution-NonCommercial 4.0 International License...}} \\
\bottomrule
\end{tabular}%
}

\resizebox{\textwidth}{!}{%
\begin{tabular}{p{3.5cm} p{8cm} p{4cm}}
\toprule
\textbf{Model} & \textbf{Output} & \textbf{Summary} \\
\midrule
\textbf{LawGPT} & ``The CC BY-NC 4.0 license allows for the use of data, but there are restrictions. You may need to attribute the original creator.'' & \textit{Vague response}, does not clarify the key commercial restrictions. \\
\midrule
\textbf{ChatGPT-4} & ``The CC BY-NC 4.0 license prohibits commercial use. You can use the dataset for non-commercial purposes, but if you plan to make money, it may violate the terms.'' & \textit{Basic response}, mentions the prohibition but lacks further details. \\
\midrule
\textbf{Qwen-1.5} & ``Under CC BY-NC 4.0, you can use the dataset as long as it is not for commercial purposes. Commercial use may be restricted.'' & \textit{Incomplete answer}, does not fully explain consequences or alternatives for commercial usage. \\
\midrule
\textbf{LicenseGPT} & ``You cannot use a dataset licensed under CC BY-NC 4.0 in a commercial project without violating the terms. The 'NC' stands for 'NonCommercial', which explicitly restricts use for any commercial purposes, including activities that involve financial gain. However, you can use the dataset for research or educational purposes. If you still want to use the dataset for commercial purposes, you would need to negotiate a separate commercial license with the rights holder. Be sure to provide proper attribution regardless of the usage type, as required by the 'BY' clause of the license.'' & \textit{Detailed analysis}, clarifies restrictions and offers a solution through licensing negotiation. \\
\bottomrule
\end{tabular}%
}
\end{table*}

\smallskip\noindent\textbf{\textit{RQ3: Do software IP lawyers find LicenseGPT to be a useful tool for expediting dataset license compliance analysis?}}

\smallskip\noindent\textbf{Approach.} In RQ3, we evaluate the usefulness of LicenseGPT for software IP lawyers through an A/B test and a user study. We invited the lawyers involved in labeling our DL dataset to assess LicenseGPT using two methods: an A/B test and a structured interview.

For the A/B test, we collected 10 additional publicly available datasets and their associated licenses, which were not part of the original DL dataset. The lawyers were divided into two groups. One group used LicenseGPT to perform a dataset license compliance analysis, determining if the dataset's license permitted commercial usage, while the other group conducted the analysis without LicenseGPT. 

In addition to the A/B test, we conducted semi-structured interviews with the lawyers and applied thematic analysis to their responses. This allowed us to assess LicenseGPT's perceived usefulness and identify areas for improvement.

\smallskip\noindent\textit{A/B Test} The A/B test evaluated whether lawyers using LicenseGPT had the same accuracy and improved efficiency during dataset license compliance analysis. Two groups of software IP lawyers participated. In Experiment A, the first group manually annotated the datasets without using LicenseGPT, generating the ``Lawyer Review Result'' as the ground truth. In Experiment B, the second group analyzed the same dataset licenses with the assistance of LicenseGPT, though manual analysis was still involved. The two groups worked independently, and we used the ``Lawyer Review Result'' from Experiment A as the ground truth for comparison. We recorded both the agreement between their judgments and the time taken for each analysis. This experiment only compared whether both groups reached the same conclusion on the commercial usability of the datasets; we did not assess the agreement in the rationale provided by each group.

We evaluated two key aspects: (1) whether the lawyers' determination of the dataset's commercial usability in Experiment B aligned with the ground truth (i.e., PA), and (2) the average time each group took to analyze the commercial usability of the dataset licenses.

\smallskip\noindent\textit{User Study} We designed a questionnaire to capture their insights into LicenseGPT’s impact on their legal practice for the four software IP lawyers who participated in our study. 

We used a semi-structured format for the interviews, encouraging open-ended responses while guiding the conversation with specific questions for consistency. Since all the participating lawyers were from China, we conducted the discussions in Chinese. Below are the translated versions of the questions we asked.

\begin{itemize}[leftmargin=*]
    \item How useful do you find LicenseGPT for completing your dataset license compliance analysis?
    \item What challenges or limitations have you faced with traditional methods of license analysis?
    \item In what ways, if any, has LicenseGPT provided unique advantages over conventional dataset license compliance analysis techniques?
    \item Have you found LicenseGPT to provide faster and more accurate interpretations compared to traditional methods? Can you provide specific examples?
    \item Would you consider using LicenseGPT as an auxiliary tool in future license compliance reviews?
    \item What potential time savings do you foresee when using LicenseGPT alongside traditional dataset license compliance analysis methods?
\end{itemize}

We digitally recorded the feedback and took detailed notes to supplement the recordings. We then conducted a thematic analysis of the responses similar to prior studies~\cite{butler2024objectives}. After transcribing and translating the responses into English, two authors independently familiarized themselves with the data by reading through the transcripts and taking initial notes. Both authors then systematically coded the data, labeling key segments related to the research question. After independently coding, they compared and refined their codes, collaboratively identifying recurring patterns and emerging themes. Through multiple meetings, they reviewed and finalized the themes, ensuring they accurately reflected the data and addressed the research questions.

\begin{table}[H]
\fontsize{10}{12}
\centering
\caption{Impact of LicenseGPT on Lawyer's Accuracy and Efficiency}
\scriptsize
\label{tab:feedback}
\begin{tabular}{lrr}
\hline
\textbf{}                        & \textbf{PA (\%)} & \textbf{Eff. (seconds)} \\ \hline
\textbf{Software IP Lawyer (w/o LicenseGPT)}  & 100 & 108 \\
\textbf{Software IP Lawyer (with LicenseGPT)} & 100 & 6 \\ \hline
\end{tabular}
\end{table}

\smallskip\noindent\textbf{Results.~Result 5. LicenseGPT significantly reduces the time required for dataset license compliance analysis, enhancing efficiency for software IP lawyers.} In our A/B test (see Table~\ref{tab:feedback}), lawyers using LicenseGPT completed analyses in an average of 6 seconds per license, compared to 108 seconds without the tool — a 94.44\% reduction in time. This substantial decrease demonstrates LicenseGPT's ability to accelerate the legal review process. Lawyer L1 estimated, \textit{``using LicenseGPT could save me around 50\% of the time I normally spend on license compliance analysis,”} highlighting the practical efficiency gains. Similarly, Lawyer L2 noted, \textit{``I believe using LicenseGPT could save me around 40-50\% of the time.''} reinforcing the tool's potential to streamline workflows.

\smallskip\noindent\textbf{Result 6. While LicenseGPT enhances efficiency, software IP lawyers recognize the need for careful validation due to limitations in handling complex legal nuances.} Feedback from the lawyers indicated appreciation for the tool's speed and user-friendliness but also cautiousness regarding its reliability in complex cases. Lawyer L3 commented, \textit{``LicenseGPT is highly advantageous for quickly filtering through datasets... a more detailed analysis of the license can then be performed.”} suggesting its utility as an initial assessment tool. However, Lawyer L4 expressed concerns: \textit{``Given the ambiguity in law, using pure AI to assess comprehensive legal risks is too risky. I naturally would not trust AI for standalone use in this field but could consider it as a supplementary tool.”} These concerns highlights the need for human oversight, especially in nuanced scenarios.

\smallskip\noindent\textbf{Result 7. LicenseGPT is perceived as a valuable supplementary tool that can be integrated into future legal workflows.} The majority of lawyers expressed willingness to incorporate LicenseGPT into their practice as an auxiliary resource. Lawyer L1 stated, \textit{``Yes, I would. While it may not completely replace manual review in more complex cases, it is very useful for initial assessments and ensuring that I don’t overlook important sections of a license.''} Similarly, Lawyer L2 affirmed, \textit{``Absolutely. LicenseGPT has proven to be an excellent supplementary tool for my legal practice.''} These sentiments indicate that while LicenseGPT may not replace traditional methods, It would basically replace the intern/law associate who does the first pass.

\smallskip\noindent\textbf{Result 8. LicenseGPT addresses challenges faced in traditional license analysis by providing quick identification of key clauses and reducing manual effort.} Lawyers noted that traditional methods are time-consuming and prone to human error. Lawyer L1 mentioned, \textit{``The process of manually searching for key clauses in complex legal documents can be overwhelming and prone to human error.”} LicenseGPT alleviates these issues by automating the initial review, allowing lawyers to focus on complex legal reasoning.

\section{Discussion}
\label{sec:disscuss}
\smallskip\noindent\textbf{How does prompt design impact the performance of LicenseGPT in dataset license compliance analysis?}

\smallskip\noindent\textbf{Motivation.} Precision is paramount in legal tasks such as dataset license comprehension. Previous studies have shown that variations in prompt design can significantly affect the performance of language models~\cite{yu-etal-2023-exploring}. Understanding the sensitivity of LicenseGPT to different system prompts and user prompts is essential to optimize its Prediction Agreement (PA) and ensure reliable outputs in legal contexts.

\smallskip\noindent\textbf{Approach.} We created six system prompts using three approaches: custom designs (Sys\_v1 to Sys\_v3), ChatGPT-4 generated prompts based on task descriptions (Sys\_v4 and Sys\_v5), and a PromptSource-generated prompt (Sys\_v6) after we provided the problem and task details~\cite{bach2022promptsource}. Additionally, we manually designed three user prompts (User\_v1 to User\_v3) to accompany the system prompts. Each system-user prompt pair was crafted to balance specificity and flexibility, ensuring clarity without overloading the model. We tested all combinations to evaluate their impact on LicenseGPT's PA.


\begin{figure}[h]
  \centering
  \includegraphics[width=0.8\columnwidth]{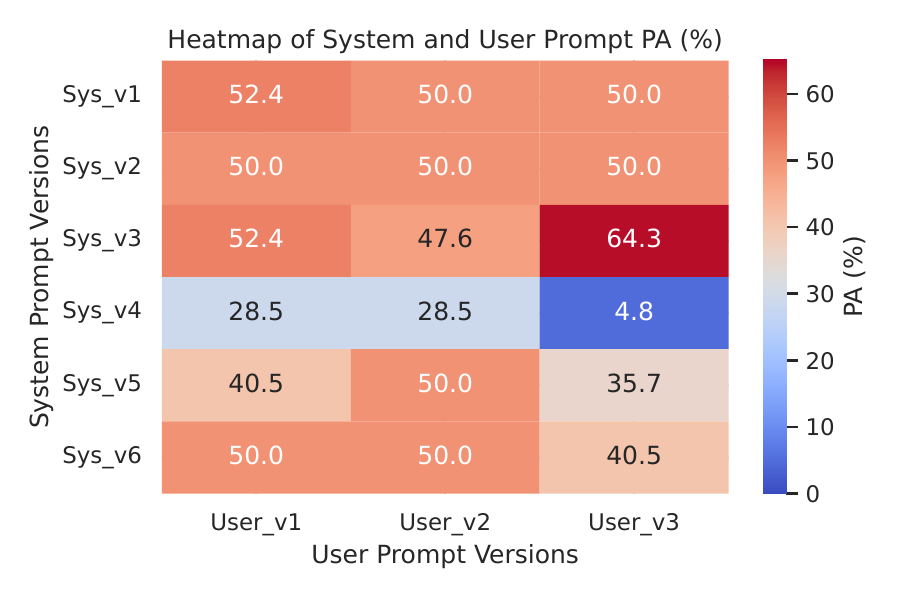}
  \caption{Heatmap of PA on studied system and user prompts}
  \label{fig:heatmap}
\end{figure}

\noindent\textbf{Results.~Result 9. Our custom prompt design significantly enhances LicenseGPT's performance in dataset license compliance analysis.} As shown in Fig~\ref{fig:heatmap}, our custom-designed system prompt \textit{Sys\_v3} combined with user prompt \textit{User\_v3} achieved the highest Prediction Agreement (PA) of 64.3\%, outperforming all other combinations. On average, our custom prompts (\textit{Sys\_v1} to \textit{Sys\_v3}) yielded higher PA scores (50.8\%, 50.0\%, and 54.8\%) compared to prompts generated by ChatGPT-4 (20.6\% and 42.1\%) and PromptSource (46.8\%). Our prompts likely outperformed others due to several key factors. First, we incorporated specific legal terminology and focused instructions that align with the nuances of dataset license compliance, emphasizing elements like ``\textit{rights and obligations analysis}'' and referencing relevant legal scenarios. Second, we explicitly defined the model's role as a software IP lawyer and clearly specified the task, reducing ambiguity and guiding the model to provide legally sound responses. Third, we structured the prompts to facilitate a step-by-step analysis, encouraging the model to thoroughly examine license terms before reaching conclusions. This tailored approach ensured the model focused on critical legal aspects, enhancing accuracy. 


\textbf{Result 10. LicenseGPT is sensitive to variations in prompt design, underscoring the importance of prompt engineering for optimal performance.}  The PA varied significantly across different prompt combinations, highlighting the model's sensitivity. For instance, using \textit{Sys\_v4} with \textit{User\_v3} resulted in a PA of only 4.8\%, whereas \textit{Sys\_v3} with the same user prompt achieved 64.3\%. This dramatic difference indicates that not just the content, but the precise phrasing and structure of prompts critically impact the model's comprehension and performance. These findings demonstrate that carefully crafted, task-specific prompts significantly improve LicenseGPT's ability to analyze dataset licenses for compliance.

\smallskip\noindent\textbf{How does the size of the instruction fine-tuning dataset affect LicenseGPT's performance?}

\smallskip\noindent\textbf{Motivation.} Labeling dataset licenses for fine-tuning LicenseGPT is both costly and time-consuming, requiring expert legal annotations from software IP lawyers. Fine-tuning even small models like LawGPT involves significant computational resources. Understanding how the size of the instruction fine-tuning dataset impacts LicenseGPT's performance is crucial to determine if investing in annotating more licenses would lead to substantial improvements and justify future efforts to expand the dataset.

\smallskip\noindent\textbf{Approach.} We conducted experiments to analyze how varying the size of the instruction fine-tuning dataset influences LicenseGPT's Prediction Agreement (PA). We fine-tuned LicenseGPT using different subsets of our DL dataset, with sizes of 100, 150, 200, 250, 300, 350, 400, and 450 licenses. For each configuration, we performed 10-fold cross-validation, similar to our approach in RQ1 and RQ2, and plotted the median PA for each data size.

\begin{table}[ht]
\fontsize{10}{12}
\centering
\scriptsize
\caption{Impact of Fine-tuning Data Size on PA}
\label{tab:DatasizeandPA}
\begin{tabular}{@{}lrrrrrrrr@{}}
\toprule
\textbf{Data Size}      & 100   & 150   & 200   & 250   & 300   & 350   & 400   & 450   \\ \midrule
\textbf{Avg PA (\%)} & 39.3  & 42.7  & 44.4  & 52.8  & 56.1  & 60.7  & 62.1  & 64.3  \\ \bottomrule
\end{tabular}
\end{table}

~\noindent\textbf{Results.~Result 11. Increasing the fine-tuning dataset size improves LicenseGPT's performance, but with diminishing returns as the size grows, and larger base models may offer further potential.} As shown in Table~\ref{tab:DatasizeandPA}, LicenseGPT's median PA rises from 39.3\% to 64.3\% as the fine-tuning dataset expands from 100 to 450 licenses. The most notable improvement occurs when increasing the dataset from 100 to 250 licenses, where PA climbs from 39.3\% to 52.8\%. Beyond 300 licenses, the performance gains taper off, suggesting diminishing returns with larger datasets. This trend indicates that expanding the dataset could further boost accuracy, though we are constrained by the costs and effort involved in legal annotations. Moreover, fine-tuning larger pre-trained models presents an additional avenue for improvement, offering the potential for more accurate and robust performance in dataset license compliance analysis.

\section{Threats to Validity}
\label{Sec:Threat}
\smallskip\noindent\textbf{Internal Validity}  
Our study may face biases in dataset labeling and FM fine-tuning. Although four software IP lawyers labeled the licenses and reached consensus, subjective interpretations and ambiguous license terms could have affected the labels and training data quality. Manual verification of FM outputs, despite care, introduces potential human error, and the design of system and user prompts may have biased responses.

\smallskip\noindent\textbf{External Validity}  
The scope of our dataset and models limits the generalizability of our findings. With 500 licenses from GitHub and Hugging Face, we may not capture all license types used in AI development. Excluding expired or unreadable licenses may introduce bias. Despite these limitations, we show that existing legal FMs are not well-suited for license compliance, but fine-tuning models with domain-specific data, as demonstrated with LicenseGPT, offers a practical solution when legal expertise is limited.

\smallskip\noindent\textbf{Interview Validity} Our user study relied on semi-structured interviews with four software IP lawyers, all located in a single country. This relatively small and geographically uniform sample may limit the diversity of perspectives and reduce the generalizability of the qualitative findings. Although guidelines often recommend 10–20 participants for robust qualitative inquiries, our smaller sample aimed to gather initial insights into LicenseGPT’s practicality. In future work, recruiting a broader range of participants—across different jurisdictions and legal systems—would yield a more inclusive view of how LicenseGPT performs under varied legal frameworks and cultural contexts.

\smallskip\noindent\textbf{Construct Validity}  
Our evaluation metrics may not fully capture the complexity of dataset license compliance. Future research should explore additional metrics. We used zero-shot settings for accessibility, though few-shot learning and fine-tuning may unlock more potential. Advanced prompt techniques like Chain-of-Thought~\cite{cot_finetuning} and Self-reflection~\cite{self_reflection} could improve FM performance. Lastly, our chosen models may not represent the entire spectrum of legal FMs, so evaluating more models could lead to different insights.

\section{Conclusion and the road ahead}
\label{sec:conclusion}
Legal compliance is a critical non-functional requirement in the AI software engineering lifecycle that directly impacts software quality. Ignoring legal issues, like copyright infringement or contract violations, undermines the reliability of AI-powered software. Our study emphasizes the need to integrate legal expertise into development, especially as laws vary across regions. While software IP lawyers are vital, reducing the effort required for license compliance analysis is essential. Tools like our LicenseGPT can streamline this process, though human oversight remains crucial.

We recommend the following for software engineers and IP lawyers: \textbf{(1)} Use LicenseGPT to simplify license compliance; \textbf{(2)} Ensure human oversight in legal evaluations; \textbf{(3)} Document compliance efforts for due diligence.

\textbf{Although LicenseGPT is designed to support Software IP lawyers, it also benefits software engineers who often perform initial license checks.} By providing timely, accurate guidance on dataset constraints, LicenseGPT enables developers to make informed decisions early in the project lifecycle, reducing the need for costly rework and fostering more streamlined collaboration with legal teams.

However, dataset license compliance analysis is a complex problem, and our work addresses only part of it. For seamless integration of dataset license compliance into AI software engineering lifecycle, we highlight these immediate challenges:

\begin{itemize}[leftmargin=*]
\item \textbf{Challenge 1: Identifying and Analyzing All Associated Licenses.} Datasets often aggregate data from various sources, each with its own license. Analyzing these, especially when licenses conflict, is difficult~\cite{Gopi2021can}. Developing tools to collect and assess all licenses is urgently needed.

\item \textbf{Challenge 2: Lack of Standardized License Metadata.} Current documentation standards like Datasheets, Factsheets, and Model Cards~\cite{gebru2021datasheets,arnold2019factsheets,mitchell2019modelcards} lack the necessary details for license compliance. The SPDX 3.0 Dataset Profile~\cite{SPDXDatasetProfile2023} is a promising start but requires improved fields for data sources. Platforms like Hugging Face should adopt standardized metadata to clarify rights and obligations across the supply chain.

\item \textbf{Challenge 3: Extending Compliance to AI Models.} Evaluating AI model licenses alongside their training datasets' licenses is essential. Automating this requires tools capable of tracking and interpreting licenses throughout the model development pipeline. Standardizing this process and incorporating it into frameworks like OpenChain~\cite{OpenChainProject} is pivotal. Researchers must formalize the compliance process and actively participate in standardization initiatives, such as the SPDX AI~\cite{SPDXAI2024} and OpenChain AI working groups~\cite{OpenChainAIWorkshop2024}, to ensure the latest research informs these standards and is translated into practice.
\end{itemize}

By addressing these challenges and adopting our recommendations, we can embed legal compliance into AI development, improving software quality and reducing legal risks for all stakeholders.





\section*{Disclaimer}
Any opinions, findings, conclusions, or recommendations expressed in this material are those of
the author(s) and do not reflect the views of our company (will be updated during camera ready). ChatGPT-4o was used only for copy-editing and table formatting in compliance with IEEE and ACM policies on AI use in publications.



%

\bibliographystyle{IEEEtranS}
\balance
\scriptsize
\bibliography{reference}

@misc{osi,
  title={Open Source Initiative},
  note={Available at: \url{https://opensource.org/licenses}},
  year={2024}
}

@misc{github_guide,
  title={GitHub Licensing Guide},
  note={\url{https://docs.github.com/en/repositories/managing-your-repositorys-settings-and-features/customizing-your-repository/licensing-a-repository}},
  year={2024}
}

@misc{tldrlegal,
  title={TLDRLegal: Understand Open Source Licenses},
  note={Available at: \url{https://www.tldrlegal.com/}},
  year={2024}
}

@article{lin2024legal,
  title={Legal Documents Drafting with Fine-Tuned Pre-Trained Large Language Model},
  author={Lin, Chun-Hsien and Cheng, Pu-Jen},
  journal={arXiv preprint arXiv:2406.04202},
  year={2024}
}

@article{huang2023lawyerLLaMA,
  title={Lawyer LLaMA Technical Report},
  author={Huang, Quzhe and Tao, Mingxu and An, Zhenwei and Zhang, Chen and Jiang, Cong and Chen, Zhibin and Wu, Zirui and Feng, Yansong},
  journal={arXiv preprint arXiv:2305.15062},
  year={2023}
}

@article{zhao2023survey,
  title={A survey of large language models},
  author={Zhao, Wayne Xin and Zhou, Kun and Li, Junyi and Tang, Tianyi and Wang, Xiaolei and Hou, Yupeng and Min, Yingqian and Zhang, Beichen and Zhang, Junjie and Dong, Zican and others},
  journal={arXiv preprint arXiv:2303.18223},
  year={2023}
}

@article{cui2023chatlaw,
  title={Chatlaw: Open-source legal large language model with integrated external knowledge bases},
  author={Cui, Jiaxi and Li, Zongjian and Yan, Yang and Chen, Bohua and Yuan, Li},
  journal={arXiv preprint arXiv:2306.16092},
  year={2023}
}

@misc{songLawGPT,
  author = {Song Pengxiao, Zhou Zhi and cainiao},
  title = {LaWGPT: Chinese-Llama tuned with Chinese Legal knowledge},
  year = {2023},
  url = {https://github.com/pengxiao-song/LaWGPT}
}

@misc{HanFei,
  author = {Jibao Wen and Wanwei He},
  title = {HanFei},
  year = {2023},
  url = {https://github.com/siat-nlp/HanFei}
}

@misc{LawGPTzh,
  author = {LiuHongcheng,LiaoYusheng,MengYutong and WangYuhao},
  title = {LawGPT:Chinese Legal Model},
  year = {2023},
  url = {https://github.com/LiuHC0428/LAW_GPT}
}

@misc{fuzi.mingcha,
  author = {Shiguang Wu, Zhongkun Liu, Zhen Zhang and others},
  title = {fuzi.mingcha},
  year = {2023},
  url = {https://github.com/irlab-sdu/fuzi.mingcha}
}

@online{LexiLaw,
  title = {LexiLaw},
  url = {https://github.com/CSHaitao/LexiLaw}
}

@online{zhihai,
  title = {wisdomInterrogatory},
  url = {https://github.com/zhihaiLLM/wisdomInterrogatory}
}

@article{touvron2023llama,
  title={Llama: Open and efficient foundation language models},
  author={Touvron, Hugo and Lavril, Thibaut and Izacard, Gautier and Martinet, Xavier and Lachaux, Marie-Anne and Lacroix, Timoth{\'e}e and Rozi{\`e}re, Baptiste and Goyal, Naman and Hambro, Eric and Azhar, Faisal and others},
  journal={arXiv preprint arXiv:2302.13971},
  year={2023}
}

@article{cui2023chinesellama,
  title={Efficient and effective text encoding for chinese llama and alpaca},
  author={Cui, Yiming and Yang, Ziqing and Yao, Xin},
  journal={arXiv preprint arXiv:2304.08177},
  year={2023}
}

@article{zeng2022glm,
  title={Glm-130b: An open bilingual pre-trained model},
  author={Zeng, Aohan and Liu, Xiao and Du, Zhengxiao and Wang, Zihan and Lai, Hanyu and Ding, Ming and Yang, Zhuoyi and Xu, Yifan and Zheng, Wendi and Xia, Xiao and others},
  journal={arXiv preprint arXiv:2210.02414},
  year={2022}
}

@inproceedings{du2022glm,
  title={GLM: General Language Model Pretraining with Autoregressive Blank Infilling},
  author={Du, Zhengxiao and Qian, Yujie and Liu, Xiao and Ding, Ming and Qiu, Jiezhong and Yang, Zhilin and Tang, Jie},
  booktitle={Proceedings of the 60th Annual Meeting of the Association for Computational Linguistics (Volume 1: Long Papers)},
  pages={320--335},
  year={2022}
}

@article{Gopi2021can,
  title={Can I use this publicly available dataset to build commercial AI software? most likely not},
  author={Rajbahadur, Gopi Krishnan and Tuck, Erika and Zi, Li and Wei, Zhang and Lin, Dayi and Chen, Boyuan and Jiang, Zhen Ming and German, Daniel Morales},
  journal={CoRR, abs/2111.02374},
  pages={1--1},
  year={2021}
}

@misc{bach2022promptsource,
      title={PromptSource: An Integrated Development Environment and Repository for Natural Language Prompts},
      author={Stephen H. Bach and others},
      year={2022},
      eprint={2202.01279},
      archivePrefix={arXiv},
      primaryClass={cs.LG}
}

@article{brown2020language,
  title={Language models are few-shot learners},
  author={Brown, Tom and Mann, Benjamin and Ryder, Nick and Subbiah, Melanie and Kaplan, Jared D and Dhariwal, Prafulla and Neelakantan, Arvind and Shyam, Pranav and Sastry, Girish and Askell, Amanda and others},
  journal={Advances in neural information processing systems},
  volume={33},
  pages={1877--1901},
  year={2020}
}

@article{zhang2023adaptiveLoRA,
  title={Adaptive budget allocation for parameter-efficient fine-tuning},
  author={Zhang, Qingru and Chen, Minshuo and Bukharin, Alexander and He, Pengcheng and Cheng, Yu and Chen, Weizhu and Zhao, Tuo},
  journal={arXiv preprint arXiv:2303.10512},
  year={2023}
}

@article{HarvardGazette2023OpenAI,
  title = {Key Issues in Writers' Case Against OpenAI Explained},
  author = {{Harvard Gazette}},
  year = {2023},
  month = sep,
  url = {https://news.harvard.edu/gazette/story/2023/09/key-issues-in-writers-case-against-openai-explained/},
}

@article{VOANews2023NYTimesSues,
  title = {NY Times Sues OpenAI, Microsoft for Allegedly Infringing Copyrighted Work},
  author = {{VOA News}},
  year = {2023},
  url = {https://www.voanews.com/a/ny-times-sues-openai-microsoft-for-allegedly-infringing-copyrighted-work/7414394.html},
}

@article{benjamin2019towards,
  title={Towards standardization of data licenses: The montreal data license},
  author={Benjamin, Misha and Gagnon, Paul and Rostamzadeh, Negar and Pal, Chris and Bengio, Yoshua and Shee, Alex},
  journal={arXiv preprint arXiv:1903.12262},
  year={2019}
}

@article{16-german2012method,
  title={A method for open source license compliance of java applications},
  author={German, Daniel and Di Penta, Massimiliano},
  journal={IEEE software},
  volume={29},
  number={3},
  pages={58--63},
  year={2012},
  publisher={IEEE}
}

@inproceedings{17-german2009license,
  title={License integration patterns: Addressing license mismatches in component-based development},
  author={German, Daniel M and Hassan, Ahmed E},
  booktitle={2009 IEEE 31st international conference on software engineering},
  pages={188--198},
  year={2009},
  organization={IEEE}
}

@article{27-kapitsaki2017automating,
  title={Automating the license compatibility process in open source software with SPDX},
  author={Kapitsaki, Georgia M and Kramer, Frederik and Tselikas, Nikolaos D},
  journal={Journal of systems and software},
  volume={131},
  pages={386--401},
  year={2017},
  publisher={Elsevier}
}

@inproceedings{51-van2014tracing,
  title={Tracing software build processes to uncover license compliance inconsistencies},
  author={Van Der Burg, Sander and Dolstra, Eelco and McIntosh, Shane and Davies, Julius and German, Daniel M and Hemel, Armijn},
  booktitle={Proceedings of the 29th ACM/IEEE international conference on Automated software engineering},
  pages={731--742},
  year={2014}
}

@inproceedings{56-zhang2010automatic,
  title={Automatic checking of license compliance},
  author={Zhang, Hongyu and Shi, Bei and Zhang, Lu},
  booktitle={2010 IEEE International Conference on Software Maintenance},
  pages={1--3},
  year={2010},
  organization={IEEE}
}

@article{barclay2019towards,
  title={Towards traceability in data ecosystems using a bill of materials model},
  author={Barclay, Iain and Preece, Alun and Taylor, Ian and Verma, Dinesh},
  journal={arXiv preprint arXiv:1904.04253},
  year={2019}
}

@article{szpyt2020RAIL,
  title={Responsible AI licenses-a real alternative to generally applicable laws?},
  author={Szpyt, Kamil},
  journal={Revista Ib{\'e}rica do Direito},
  volume={1},
  number={2},
  pages={178--186},
  year={2020}
}

@inproceedings{amershi2019software,
  title={Software engineering for machine learning: A case study},
  author={Amershi, Saleema and Begel, Andrew and Bird, Christian and DeLine, Robert and Gall, Harald and Kamar, Ece and Nagappan, Nachiappan and Nushi, Besmira and Zimmermann, Thomas},
  booktitle={2019 IEEE/ACM 41st International Conference on Software Engineering: Software Engineering in Practice (ICSE-SEIP)},
  pages={291--300},
  year={2019},
  organization={IEEE}
}

@misc{githubcopilot2021,
  author = {{GitHub}},
  title = {GitHub Copilot: Your AI pair programmer},
  year = 2021,
  howpublished = {\url{https://copilot.github.com}},
  note = {Accessed: 2024-07-03}
}

@article{longpre2023data,
  title={The data provenance initiative: A large scale audit of dataset licensing \& attribution in ai},
  author={Longpre, Shayne and Mahari, Robert and Chen, Anthony and Obeng-Marnu, Naana and Sileo, Damien and Brannon, William and Muennighoff, Niklas and Khazam, Nathan and Kabbara, Jad and Perisetla, Kartik and others},
  journal={arXiv preprint arXiv:2310.16787},
  year={2023}
}

@misc{canadacopyright, 
    title   ={A guide to copyright},
    url = {https://laws-lois.justice.gc.ca/eng/acts/c-42/page-9.html},
    journal ={Canadian Intellectual Property Office}, 
    author  ={Government of Canada},
    note    ={[Last visited on 09-25-2024]},
    year    ={2021}
}

@article{peng2021mitigating,
  title={Mitigating dataset harms requires stewardship: Lessons from 1000 papers},
  author={Peng, Kenny and Mathur, Arunesh and Narayanan, Arvind},
  journal={arXiv preprint arXiv:2108.02922},
  year={2021}
}

@misc{lawsuit, 
    title   ={US Court of Appeals, New York, 13-4829, 2015.},
    author  ={The Authors Guild v. Google},
    year    ={2015}
}

@misc{USFairUse, 
    title   ={More Information on Fair Use},
    url = {https://www.copyright.gov/fair-use/more-info.html},
    journal ={US Copyright Office}, 
    author  ={US Copyright Office},
    note    ={[Last visited on 09-25-2024]},
    year    ={2021}
}

@misc{CanadaFairDealing, 
    author  = {{Government of Canada}},
    title   = {Infringement of Copyright and Moral Rights and Exceptions to Infringement (continued)},
    year    = {2021},
    url     = {https://laws-lois.justice.gc.ca/eng/acts/c-42/page-9.html},
    note    = {[Last visited on 09-25-2024]},
    journal = {Justice Laws Website}
}

@article{triaille2014study,
  title={Study on the legal framework of text and data mining (TDM)},
  author={Triaille, Jean-Paul and others},
  journal={European Union Studies KM-03-13-42},
  year={2014},
  publisher={European Union}
}

@inproceedings{butler2024objectives,
  title={Objectives and Key Results in Software Teams: Challenges, Opportunities and Impact on Development},
  author={Butler, Jenna L and Zimmermann, Thomas and Bird, Christian},
  booktitle={Proceedings of the 46th International Conference on Software Engineering: Software Engineering in Practice},
  pages={358--368},
  year={2024}
}

@article{ingolfo2013arguing,
  title={Arguing regulatory compliance of software requirements},
  author={Ingolfo, Silvia and Siena, Alberto and Mylopoulos, John and Susi, Angelo and Perini, Anna},
  journal={Data \& Knowledge Engineering},
  volume={87},
  pages={279--296},
  year={2013},
  publisher={Elsevier}
}

@article{breaux2008analyzing,
  title={Analyzing Regulatory Rules for Privacy and Security Requirements},
  author={Breaux and others},
  journal={IEEE Transactions on Software Engineering},
  year={2008}
}

@inproceedings{kiyavitskaya2008automating,
  title={Automating the Extraction of Rights and Obligations for Regulatory Compliance},
  author={Kiyavitskaya, Nadzeya and Zeni, Nicola and Breaux, Travis D and Ant{\\'o}n, Annie I and Cordy, James R and Mich, Luisa and Mylopoulos, John},
  booktitle={Proceedings of the 27th International Conference on Conceptual Modeling, Barcelona, Spain, October 20-24},
  year={2008}
}

@article{zeni2015gaiust,
  title={GaiusT: Supporting the Extraction of Rights and Obligations for Regulatory Compliance},
  author={Zeni, Nicola and Kiyavitskaya, Nadzeya and Mich, Luisa and Cordy, James R and Mylopoulos, John},
  journal={Requirements Engineering},
  volume={20},
  pages={1--22},
  year={2015},
  publisher={Springer}
}

@misc{pddl2018,
  title = {Open Data Commons Public Domain Dedication and License (PDDL)},
  year = {2018},
  url = {https://opendatacommons.org/licenses/pddl/},
  note = {Open Data Commons License}
}

@misc{ccby2013,
  title = {Creative Commons Attribution License (CC BY)},
  author = {Creative Commons},
  year = {2013},
  url = {https://creativecommons.org/licenses/by/4.0/},
  note = {Creative Commons License}
}

@case{jacobsen2008,
  title     = {Jacobsen v. Katzer},
  volume    = {535},
  reporter  = {F.3d},
  pages     = {1373--1381},
  court     = {United States Court of Appeals for the Federal Circuit},
  year      = {2008},
}

@article{btlj2015,
  author  = {Bradley M. Kuhn and Karen M. Sandler},
  title   = {Enforcing the GPL and Open Source Software Licenses in the US After Jacobsen v. Katzer},
  journal = {Berkeley Technology Law Journal},
  volume  = {27},
  year    = {2012},
  pages   = {231--274},
}

@online{wsgr2017,
  author       = {Wilson Sonsini Goodrich \& Rosati},
  title        = {Open Source Software: Risks, Compliance, and Best Practices},
  year         = {2017},
  url          = {https://www.wsgr.com/en/insights/open-source-software-risks-compliance-and-best-practices.html},
}

@article{jaeger2009fossology,
  title     = {FOSSology: The Open Source License Compliance Tool},
  author    = {Jaeger, Michael C. and Herzwurm, Gregor J. and B{\"o}hm, J{\"u}rgen},
  journal   = {International Free and Open Source Software Law Review},
  volume    = {1},
  number    = {2},
  pages     = {153--171},
  year      = {2009},
}

@inproceedings{hansen2010fossology,
  title     = {FOSSology: A License Compliance Tool},
  author    = {Hansen, Frank and Becker, Bernhard and Chamas, Christophe and Germain, Philippe},
  booktitle = {IFIP International Conference on Open Source Systems},
  pages     = {47--62},
  year      = {2010},
  publisher = {Springer},
}

@online{blackduck_website,
  author       = {Black Duck Software},
  title        = {Open Source Security and License Compliance Management},
  year         = {2023},
  url          = {https://www.blackducksoftware.com},
}

@misc{huggingface,
  title = {Hugging Face},
  year = {2023},
  url = {https://huggingface.co/},
  note = {Accessed: 2024-10-02}
}

@misc{googlecloud,
  title = {Google Cloud},
  year = {2023},
  url = {https://cloud.google.com/},
  note = {Accessed: 2024-10-02}
}

@misc{kaggle,
  title = {Kaggle},
  year = {2023},
  url = {https://www.kaggle.com/},
  note = {Accessed: 2024-10-02}
}

@misc{zenodo,
  title = {Zenodo},
  year = {2023},
  url = {https://zenodo.org/},
  note = {Accessed: 2024-10-02}
}

@misc{github,
  title = {GitHub},
  year = {2023},
  url = {https://github.com/},
  note = {Accessed: 2024-10-02}
}

@misc{gitlab,
  title = {GitLab},
  year = {2023},
  url = {https://gitlab.com/},
  note = {Accessed: 2024-10-02}
}

@misc{microsoftazure,
  title = {Microsoft Azure},
  year = {2023},
  url = {https://azure.microsoft.com/},
  note = {Accessed: 2024-10-02}
}

@misc{datahub,
  title = {DataHub},
  year = {2023},
  url = {https://datahub.io/},
  note = {Accessed: 2024-10-02}
}

@misc{amazons3,
  title = {Amazon S3},
  year = {2023},
  url = {https://aws.amazon.com/s3/},
  note = {Accessed: 2024-10-02}
}

@misc{figshare,
  title = {Figshare},
  year = {2023},
  url = {https://figshare.com/},
  note = {Accessed: 2024-10-02}
}

@misc{opendataology,
  title = {OpenDataology},
  year = {2023},
  url = {http://www.opendataology.com:30800/\#/dataSetAll},
  note = {Accessed: 2024-10-02}
}

@inproceedings{LegalBench,
 author = {Guha, Neel and Nyarko and others},
 booktitle = {Advances in Neural Information Processing Systems},
 editor = {A. Oh and T. Naumann and A. Globerson and K. Saenko and M. Hardt and S. Levine},
 pages = {44123--44279},
 publisher = {Curran Associates, Inc.},
 title = {LegalBench: A Collaboratively Built Benchmark for Measuring Legal Reasoning in Large Language Models},
 url = {https://proceedings.neurips.cc/paper_files/paper/2023/file/89e44582fd28ddfea1ea4dcb0ebbf4b0-Paper-Datasets_and_Benchmarks.pdf},
 volume = {36},
 year = {2023}
}

@article{patil2024review,
  title={A review of current trends, techniques, and challenges in large language models (llms)},
  author={Patil, Rajvardhan and Gudivada, Venkat},
  journal={Applied Sciences},
  volume={14},
  number={5},
  pages={2074},
  year={2024},
  publisher={MDPI}
}

@misc{getindata2023,
  author = {{GetInData}},
  title = {Large Language Models: The Legal Aspects of Licensing for Commercial Purposes},
  year = {2023},
  url = {https://getindata.com/blog/large-language-models-legal-aspects-licensing-commercial-purposes/},
  note = {Accessed: 2024-10-02}
}

@article{yadav2023scaling,
  title={Scaling Evidence-based Instructional Design Expertise through Large Language Models},
  author={Yadav, Gautam},
  journal={arXiv preprint arXiv:2306.01006},
  year={2023}
}

@article{ScottKnottESD,
  title={The Impact of Automated Parameter Optimization for Defect Prediction Models},
  author={Tantithamthavorn, Chakkrit and McIntosh, Shane and Hassan, Ahmed E. and Matsumoto, Kenichi},
  journal={IEEE Transactions on Software Engineering},
  year={2018},
  publisher={IEEE},
  doi={10.1109/TSE.2018.2794977}
}

@book{cohen1988statistical,
  title={Statistical Power Analysis for the Behavioral Sciences},
  author={Cohen, Jacob},
  year={1988},
  publisher={Lawrence Erlbaum Associates},
  address={Hillsdale, NJ},
  edition={2nd}
}

@article{OpenAI2023,
  title={ChatGPT-4 Performance on Legal Benchmarks: Evaluating its Applicability for Specialized Tasks},
  author={Bommarito, Michael and Katz, Daniel},
  journal={Artificial Intelligence and Law},
  year={2023},
  url={https://link.springer.com/article/10.1007/s10506-023-09356-y}
}

@misc{openai2023gpt4,
  title = {GPT-4},
  author = {OpenAI},
  year = {2023},
  howpublished = {\url{https://openai.com/gpt-4}},
  note = {Accessed: 2024-10-04}
}

@misc{meta2023llama,
  title = {LLaMA-2: Open and Efficient Foundation Language Models},
  author = {Meta AI},
  year = {2023},
  howpublished = {\url{https://ai.meta.com/llama}},
  note = {Accessed: 2024-10-04}
}

@misc{alibaba2023qwen,
  title = {Qwen: Open-Source Pretrained Large-Scale Language Model},
  author = {Alibaba DAMO Academy},
  year = {2023},
  howpublished = {\url{https://modelscope.cn/models/damo}},
  note = {Accessed: 2024-10-04}
}

@misc{huggingface_zeroshot,
  author       = {Hugging Face},
  title        = {What is Zero-Shot Classification?},
  year         = {2023},
  url          = {https://huggingface.co/docs/transformers/main/en/task_summary#zero-shot-classification},
  note         = {Accessed: 2024-10-05}
}

@misc{neptuneai_zeroshot,
  author       = {Neptune AI},
  title        = {Zero-shot learning: What, How, and Why it Matters for NLP},
  year         = {2023},
  url          = {https://neptune.ai/blog/zero-shot-learning},
  note         = {Accessed: 2024-10-05}
}

@inproceedings{yu-etal-2023-exploring,
    title = "Exploring the Effectiveness of Prompt Engineering for Legal Reasoning Tasks",
    author = "Yu, Fangyi  and
      Quartey, Lee  and
      Schilder, Frank",
    editor = "Rogers, Anna  and
      Boyd-Graber, Jordan  and
      Okazaki, Naoaki",
    booktitle = "Findings of the Association for Computational Linguistics: ACL 2023",
    month = jul,
    year = "2023",
    address = "Toronto, Canada",
    publisher = "Association for Computational Linguistics",
    url = "https://aclanthology.org/2023.findings-acl.858",
    doi = "10.18653/v1/2023.findings-acl.858",
    pages = "13582--13596",
}

@article{cot_finetuning,
  title={The CoT Collection: Improving Zero-shot and Few-shot Learning of Language Models via Chain-of-Thought Fine-Tuning},
  author={Zelikman, Eytan and others},
  journal={arXiv preprint arXiv:2305.14045},
  year={2023}
}

@thesis{self_reflection,
  title={Self-Reflective Chain-of-Thought Reasoning in Large Language Models},
  author={Researcher, Thesis},
  school={KTH Royal Institute of Technology},
  year={2023},
  type={Thesis}
}

@article{hassan2024rethinking,
  title={Rethinking Software Engineering in the Foundation Model Era: From Task-Driven AI Copilots to Goal-Driven AI Pair Programmers},
  author={Hassan, Ahmed E and Oliva, Gustavo A and Lin, Dayi and Chen, Boyuan and Ming, Zhen and others},
  journal={arXiv preprint arXiv:2404.10225},
  year={2024}
}

@article{munir2021OSPO,
  title={The Rise of Open Source Program Office},
  author={Munir, Hussan and Mols, Carl-Erik},
  journal={IT Professional},
  volume={23},
  number={1},
  pages={27--33},
  year={2021},
  publisher={IEEE}
}

@misc{OpenChainAIWorkshop2024,
  author       = {{OpenChain Project}},
  title        = {{OpenChain AI Study Group Monthly Workshop for North America and Europe: Full Recording}},
  howpublished = {\url{https://openchainproject.org/news/2024/04/09/openchain-ai-study-group-monthly-workshop-for-north-/america-and-europe-2024-04-02-full-recording}},
  note         = {Last accessed: October 10, 2024},
  year         = {2024}
}

@misc{OpenChainProject,
  organization = {OpenChain Project},
  title        = {OpenChain Project},
  howpublished = {\url{https://openchainproject.org/}},
  note         = {Accessed: 2024-10-10},
  year         = {2024}
}

@misc{LicenseGPT,
  author       = {{OpenDataology}},
  title        = {{LicenseGPT}},
  howpublished = {\url{https://github.com/OpenDataology/LicenseGPT}},
  note         = {GitHub repository, Last accessed: 2024-10-11},
  year         = {2024}
}

@article{gebru2021datasheets,
  title={Datasheets for datasets},
  author={Gebru, Timnit and Morgenstern, Jamie and Vecchione, Briana and Vaughan, Jennifer Wortman and Wallach, Hanna and Iii, Hal Daum{\'e} and Crawford, Kate},
  journal={Communications of the ACM},
  volume={64},
  number={12},
  pages={86--92},
  year={2021},
  publisher={ACM New York, NY, USA}
}

@inproceedings{mitchell2019modelcards,
  title={Model cards for model reporting},
  author={Mitchell, Margaret and Wu, Simone and Zaldivar, Andrew and Barnes, Parker and Vasserman, Lucy and Hutchinson, Ben and Spitzer, Elena and Raji, Inioluwa Deborah and Gebru, Timnit},
  booktitle={Proceedings of the conference on fairness, accountability, and transparency},
  pages={220--229},
  year={2019}
}

@article{arnold2019factsheets,
  title={FactSheets: Increasing trust in AI services through supplier's declarations of conformity},
  author={Arnold, Matthew and Bellamy, Rachel KE and Hind, Michael and Houde, Stephanie and Mehta, Sameep and Mojsilovi{\'c}, Aleksandra and Nair, Ravi and Ramamurthy, K Natesan and Olteanu, Alexandra and Piorkowski, David and others},
  journal={IBM Journal of Research and Development},
  volume={63},
  number={4/5},
  pages={6--1},
  year={2019},
  publisher={IBM}
}

@misc{SPDXDatasetProfile2023,
  title        = {{SPDX 3.0 Dataset Profile}},
  year         = {2023},
  url          = {https://spdx.github.io/spdx-spec/v3.0/model/Dataset/Dataset/},
  note         = {Accessed: 2024-10-11},
}

@misc{SPDXAI2024,
  title        = {{SPDX AI - Areas of Interest}},
  year         = {2024},
  url          = {https://spdx.dev/learn/areas-of-interest/ai/},
  note         = {Accessed: 2024-10-11},
}

@article{dunn1961Bonferroni,
  title={Multiple comparisons among means},
  author={Dunn, Olive Jean},
  journal={Journal of the American statistical association},
  volume={56},
  number={293},
  pages={52--64},
  year={1961},
  publisher={Taylor \& Francis}
}

@article{rupert2012Bonferroni,
  title={Simultaneous statistical inference},
  author={Rupert Jr, G and others},
  year={2012},
  publisher={Springer Science \& Business Media}
}

\end{document}